\documentclass[12pt,nohyper,letterpaper]{JHEP3}         
\usepackage{amssymb,amsfonts}

\usepackage{epsfig}

\def\be{\begin{eqnarray}}
\def\ee{\end{eqnarray}}
\newcommand{\nn}{\nonumber}
\newcommand\para{\paragraph{}}
\newcommand{\ft}[2]{{\textstyle\frac{#1}{#2}}}
\newcommand{\eqn}[1]{(\ref{#1})}

\def\Dslash{\,\,{\raise.15ex\hbox{/}\mkern-12mu D}}
\def\Dbarslash{\,\,{\raise.15ex\hbox{/}\mkern-12mu {\bar D}}}
\def\delslash{\,\,{\raise.15ex\hbox{/}\mkern-9mu \partial}}
\def\delbarslash{\,\,{\raise.15ex\hbox{/}\mkern-9mu {\bar\partial}}}
\def\pslash{\,\,{\raise.15ex\hbox{/}\mkern-9mu p}}
\def\calDslash{\,\,{\raise.15ex\hbox{/}\mkern-12mu {\cal D}}}

\newcommand\Tr{{\rm Tr}}

\newcommand{\CP}{{\bf CP}}

\title{\huge Quantum Vortex Strings: A Review}

\author{David Tong\\
Department of Applied Mathematics and Theoretical Physics, \\
Centre for Mathematical Sciences, \\
Wilberforce Road, \\
Cambridge, CB3 OBA, UK \\
\email{d.tong@damtp.cam.ac.uk} }

\abstract{The quantum worldsheet dynamics of vortex strings contains information about the 4d non-Abelian gauge theory in which the string lives. Here I tell this story. The string  worldsheet theory is typically some variant of the ${\bf CP}^{N-1}$ sigma-model, describing the orientation of the string in a $U(N)$ gauge group. Qualitative parallels between 2d sigma-models and 4d non-Abelian gauge theories have been known since the 1970s. The vortex string provides a quantitative link between the two. In 4d theories with ${\cal N}=2$ supersymmetry, the exact BPS spectrum of the worldsheet coincides with the bulk spectrum in 4d. Moreover, by tuning parameters, the ${\bf CP}^{N-1}$ sigma-model can be coaxed to flow to an interacting conformal fixed point which is related to the 4d Argyres-Douglas fixed point. For theories with ${\cal N}=1$ supersymmetry, the worldsheet theory suffers dynamical supersymmetry breaking and, more interestingly, supersymmetry restoration, in a way which captures the physics of Seiberg's quantum deformed moduli space.
\para
{}
\para
\begin{center}
{\it Adams Prize Essay}\end{center}}

\begin{document}

\section{Introduction}

In the 1970s, faced with the task of understanding strong coupling effects in QCD, theorists did what good theorists always do: they explored toy models. Surprisingly, one of the most illuminating of these toys did not live in four spacetime dimensions, but rather in two. It was the
${\bf CP}^{N-1}$ sigma-model, first introduced in \cite{eich,golo,cremmer}. This model captures many of the most important features of QCD, including asymptotic freedom, the generation of a mass gap, confinement, chiral symmetry breaking, anomalies, instantons and a large N expansion \cite{dadda1,dadda,beaut}. Yet, in stark contrast to QCD, each of these properties can be demonstrated with ease. Historically, this proved to be important. Understanding the dynamics of the ${\bf CP}^{N-1}$ model led to a resolution of the tensions between the $1/N$ expansion, instantons and the quark model \cite{beaut}, as well as to the celebrated Witten-Veneziano formula for the $\eta^\prime$ mass \cite{wv,justv}. Many of the qualitative parallels that lie between 2d sigma-models and 4d non-Abelian gauge theories are described in the early review \cite{nsvz}.

\DOUBLEFIGURE{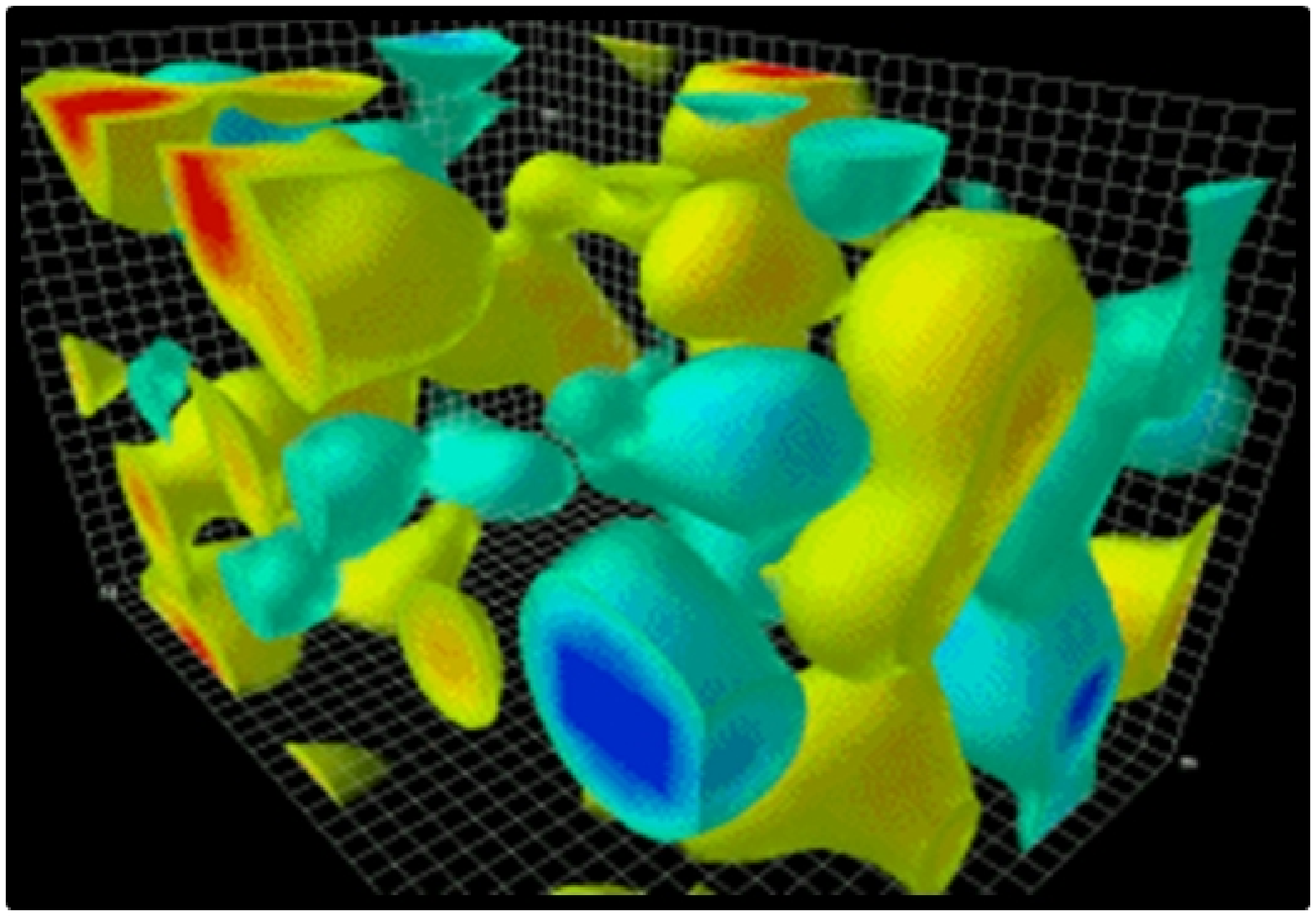,width=220pt}{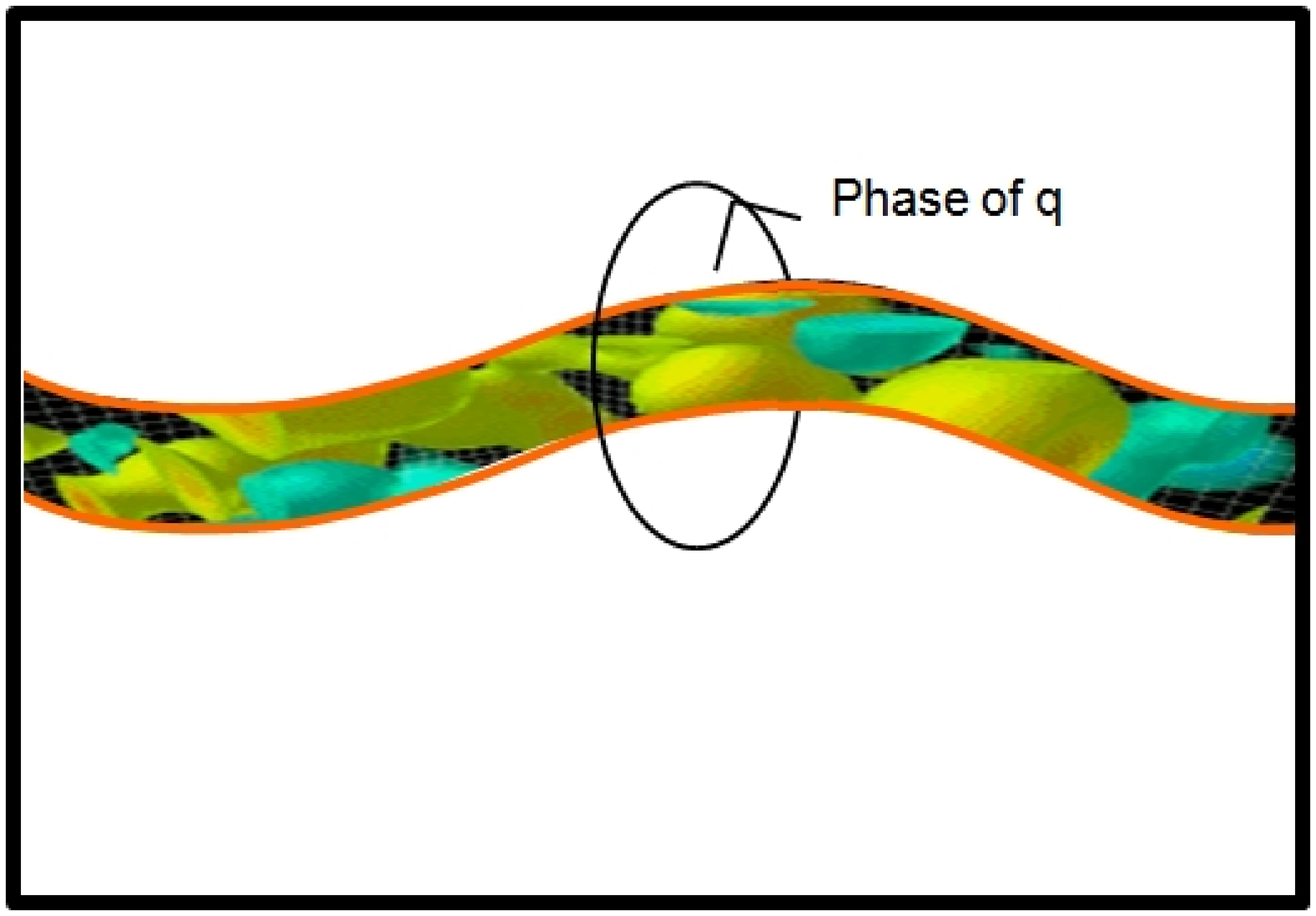,width=220pt} {The strongly coupled phase. (The graphics are taken from \cite{lein}).}{The weakly coupled Higgs phase. The core of the vortex string remains strongly coupled.}

\para
The purpose of this article is to review more recent work which provides a physical explanation for the relationship between 2d sigma-models and 4d gauge theories and also extends the parallels from the realm of the qualitative to the quantitative. Key to this approach is a map between certain 4d gauge theories and 2d non-linear sigma-models. This map is provided by solitonic vortex strings. As we will explain in detail in Section 2, vortex strings in a $U(N)$ gauge theory inherit orientational modes and their low-energy dynamics is  described by some variant of the ${\bf CP}^{N-1}$ sigma-model on the string worldsheet \cite{vib,auzzi}. In Section 3 and Section 4 we will show how the 2d dynamics of the string captures information about the quantum behaviour of the 4d gauge theory in which it lives.

\para
In the rest of this introduction, we will describe how this works in a little more detail. Our starting point is a $U(N_c)$ gauge theory, coupled to a number $N_f$ of fundamental scalar fields
$q$, where $N_f\geq N_c$. In general, the low-energy physics of interest is strongly coupled,
as represented by the wild oscillations shown in Figure 1. This is the physics that we would
like to understand.

\para
To study this system, we first deform the theory, pushing it into the Higgs phase by inducing a vacuum
expectation value for $q$. If the expectation value is made sufficiently large, and the gauge group is fully broken, then this deformed theory will be weakly coupled. All wild quantum fluctuations are suppressed, represented graphically by the boring, white background of Figure 2. At first glance, this doesn't look like a promising route to understanding the physics of the strongly coupled phase. However, the theory in the Higgs phase also admits semi-classical vortex strings, stabilized by the phase of $q$ winding in the plane transverse to the string.  While meteorological disaster movies lead us to view the core of the vortex as an oasis of tranquility, in non-Abelian gauge theories this couldn't be further from the truth. Unlike the bulk of spacetime, the core of the vortex is subject to strong coupling quantum effects. It is a wildly fluctuating region in a sea of otherwise Higgs-induced calm. Moreover, as we shall review below, the dynamics of the vortex core retains information about the original, normal state, of the 4d theory.

\para
In Section 2, we describe classical aspects of the vortex string. In particular, we review how the vortex inherits a ${\bf CP}^{N_c-1}$ moduli space of orientational modes from the $U(N_c)$ gauge group in which it lives \cite{vib,auzzi}. We also describe how magnetic monopoles appear in the theory, confined to live on the vortex string where they appear as kinks \cite{memono}. Finally we briefly sketch the intricate web of other classical BPS solitons that exist in these theories, and review some potential applications for non-Abelian vortex strings.

%
%

\para
In Section 3, we turn to vortices in 4d theories with ${\cal N}=2$ supersymmetry. Thanks to the famous work of Seiberg and Witten \cite{sw,sw2}, the low-energy dynamics and the BPS spectrum of these theories is known. We will show how this information is captured by the 2d dynamics of the vortex string \cite{sy,vstring}. In particular, following earlier work of \cite{nick,dht}, we will show that the exact quantum spectrum of BPS excitations of the string coincides with the BPS spectrum of the 4d theory. The quarks and W-bosons appear as elementary excitations of the string, while the monopoles, which are necessarily confined in the Higgs phase, appear as kinks on the vortex string. We will further see how the Argyres-Douglas points in 4d correspond to conformal field theories on the vortex worldsheet.

\para
Section 4 deals with vortices in 4d theories with ${\cal N}=1$ supersymmetry. These have ${\cal N}=(0,2)$ supersymmetry, and are referred to as ``heterotic vortex strings". Once again, the worldsheet dynamics of the string captures information about the 4d quantum dynamics. In particular, we shall show how dynamical supersymmetry breaking and, more interestingly, dynamical supersymmetry restoration on the string worldsheet corresponds to Seiberg's quantum deformation of the 4d vacuum moduli space.

\para
There have been a number of other recent reviews on the related topics of solitons in non-Abelian gauge theories \cite{tasi,mmatrix} and their quantization \cite{syreview,konrev}.

\section{Non-Abelian Vortex Strings}

We start by describing the simplest theory in which non-Abelian vortices arise. We work in four-dimensional Minkowski space, with gauge group $U(N_c)$ and $N_f$ scalars
$q_i$, $i=1,\ldots, N_f$, transforming in the fundamental representation. The Lagrangian is given by,
\be
{\cal L}_{4d}=-\frac{1}{2e^2}{\rm Tr} F_{\mu\nu}F^{\mu\nu}+\sum_{i=1}^{N_f}|{\cal D}_\mu q_i|^2
-\frac{e^2}{2}{\rm Tr}(\ \sum_{i=1}^{N_f}q_iq_i^\dagger-v^2)^2\ .
\label{theory}\ee
Here the combination $q_iq_i^\dagger$ in the final term is an $N_c\times N_c$ matrix and similarly the Higgs expectation value $v^2$ is to be thought of as multiplying the $N_c\times N_c$ unit matrix. The classical phase of the theory depends on the relative
values of $N_f$ and $N_c$. For $N_f<N_c$, an unbroken gauge group remains. In contrast, for
$N_f\geq N_c$, the gauge group is broken completely and the theory lies in the Higgs phase.
We will be interested in the Higgs phase because this is where vortices exist. For simplicity, we will choose to focus on the case $N_f=N_c\equiv N$. In this case the theory has a unique vacuum state which, up to a gauge transformation, can be written as
\be q^a_{\ i}=v\delta^a_{\ i}\ ,\ee
where $i=1,\ldots, N_f=N$ is the flavour index and $a=1,\ldots,N_c=N$ is the colour index.  In this vacuum the $U(N_c)$ gauge
group and the $SU(N_f)$ flavour group are broken to
\be
U(N_c)\times SU(N_f)\rightarrow SU(N)_{\rm diag}\ ,
\label{symmetry}\ee
a symmetry breaking pattern known as the colour-flavour locking. The theory exhibits a classical mass gap at the scale $ev$.

\para
The fact that the overall $U(1) \subset U(N_c)$ is broken in the vacuum provides the necessary topology to ensure the existence of vortex strings. These are supported by the asymptotic winding of $q$ in the $z=x^1+ix^2$ plane, transverse to the string. This winding is labeled by the integer $k \in \Pi_1(U(1))\cong {\bf Z}$. For the specific choice of scalar potential in \eqn{theory}, the vortices obey first-order differential equations of the Bogomolnyi type \cite{bog}. For $k<0$, these read
\be
(B_3)^a_{\ b}=e^2\left(\sum_{i=1}^{N_f}q^a_{\ i}q^\dagger_{ib}-v^2\delta^a_{\ b}
\right) \ \ \ ,\ \ \ \ ({\cal D}_zq_i)^a=0 \ ,
\label{vortex}\ee
with $B_\mu=\ft12\epsilon_{\mu\nu\rho}F^{\nu\rho}$, for spatial indices $\mu,\nu,\rho=1,2,3$. Solutions to these equations describe infinite, straight strings, lying in the $x_3$ direction,  with tension,
\be T = - v^2\,{\rm Tr}\int dx^1dx^2\,B_3 = 2\pi v^2|k|\ .\ee

\para
For the Abelian-Higgs model (i.e $N=1$), the equations \eqn{vortex} have long been studied and describe Abrikosov-Nielsen-Olesen \cite{abrik,no} vortices at critical coupling (at the border between type I and type II superconductivity). Although no analytic solutions to the equations are known, their existence has been demonstrated by Taubes \cite{taubes} for general $k$, and it is a simple matter to find solutions numerically. The single $|k|=1$ vortex has width $L_{\rm vort} \sim 1/ev$.

\para
The $N>1$ vortex equations \eqn{vortex} are a non-Abelian generalisation of the Abrikosov-Nielsen-Olesen vortices, which were first studied in \cite{vib,auzzi}. For a single $|k|=1$ vortex, the most general solution can be constructed by appropriately embedding an Abelian vortex into an $N\times N$ matrix. Let us call the solution to the Abelian equations $B^\star$ and $q^\star$. Then we may, for example,  embed the vortex in the upper-left hand corner of the gauge and flavour groups,
\be
B^a_{\ b}=\tiny{\left(\begin{array}{cccc} B^\star & & & \\ & 0 & & \\ & & \ddots & \\ & & & 0
\end{array}\right)}\ \ \ \ \ \ , \ \ \ \ \ \ q^a_{\ i}=\tiny{\left(\begin{array}{cccc} q^\star
&  & & \\ & v & & \\ & & \ddots & \\ & & & v\end{array}\right)}
\ .
\ee
But, of course, this is not the most general embedding.  New solutions can be generated by acting with the $SU(N)_{\rm diag}$ symmetry \eqn{symmetry} that leaves to vacuum invariant. There is an $SU(N-1)\times U(1)$ stabilizer of this group action which does not touch the vortex. Dividing by this stabilizer, we find a moduli space of solutions given by
the coset space \cite{vib,auzzi}\footnote{Strictly speaking, this argument only works in singular gauge, where the Higgs field $q$ does not wind asymptotically \cite{auzzi}. To construct the most general solution in non-singular gauge, one must work in patches over the moduli space -- see, for example, \cite{3dcs}.},
\be
SU(N)_{\rm diag}/SU(N-1)\times U(1) \cong {\bf CP}^{N-1}\ .
\label{coset}\ee
Different points in $\CP^{N-1}$ describe different orientations of the vortex in colour and flavour space.

\subsubsection*{Low Energy Dynamics}

We now look at the way these strings move. We restrict attention to an infinite, slowly-moving, almost-straight string. The string can oscillate in the two transverse directions. We describe this by promoting the center of mass coordinate $Z$, which governs the position of the vortex in the $x^1-x^2$ plane, to a dynamical field which can vary in time and along the string: $Z=Z(x^0,x^3)$. The string also contains orientational degrees of freedom and these too can vary along the string worldsheet. The associated coordinates on ${\bf CP}^{N-1}$ are therefore also promoted to dynamical fields whose oscillations give rise to classically massless modes on the string, somewhat akin to spin waves. The result is a two-dimensional non-linear sigma-model, mapping the the string worldsheet into the moduli space $\CP^{N-1}$.

\para
It is simplest to describe the ${\bf CP}^{N-1}$ sigma-model by introducing homogeneous coordinates $\varphi_i(x^0,x^3)$, $i=1\ldots, N$, together with an auxiliary $U(1)$ gauge field $u_\alpha(x^0,x^3)$ on the string worldsheet \cite{beaut}. The low-energy dynamics of the string is then given by the two-dimensional theory,
\be {\cal L}_{\rm vortex} = \ft12T|\partial_\alpha Z|^2 + |{\cal D}_\alpha\varphi_i|^2 + \lambda\left(|\varphi_i|^2-r\right)\ .\label{cpn}\ee
where $\alpha=0,3$ and ${\cal D}\varphi= \partial\varphi-iu\varphi$.  The field $\lambda$ is a Lagrange multiplier, imposing the constraint $\sum_i|\varphi_i|^2=r$. There is no kinetic term for the gauge field $u$. Instead its role is to identify configurations related by the $U(1)$ gauge action $\varphi_i\sim e^{i\alpha}\varphi_i$. These two conditions define the target space $\CP^{N-1}$,
\be \{\sum_{i=1}^N\left.|\varphi_i|^2 = r \ \right|\ \varphi_i\sim e^{i\alpha}\varphi_i \} \cong \CP^{N-1}\ .\ee
It remains to determine $r$, which governs the moment of inertia of our orientational modes. Geometrically, it is the size of the $\CP^{N-1}$ (more precisely, it is proportional to the K\"ahler class). It can be shown to be given by \cite{vib,gsy},
\be r = \frac{4\pi}{e^2}\ .\label{r}\ee
We see that when the 4d theory is weakly coupled ($e^2\ll 1$), the 2d worldsheet theory is also weakly coupled ($r\gg 1$).

\subsubsection*{Initial Comments on Quantum Effects}

Sections 3 and 4 will be devoted to understanding the quantum dynamics of vortex strings in supersymmetric theories. Here we make a few general comments.
In the absence of the Higgs mechanism, the 4d theory will flow to strong coupling at the scale $\Lambda_{4d}$. What actually happens depends on the ratio of the scales $\Lambda_{4d}$ and $ev$, the mass of the W-bosons.

\para
For $\Lambda_{4d}\gg ev$, the theory hits strong coupling before the Higgs mechanism has a chance to take effect. Here we must first solve the low-energy four-dimensional physics before we can begin to discuss vortices. Typically, we can only do this quantitatively in supersymmetric theories and we will give examples in Sections 3 and 4.

\para
In contrast, when  $ev \gg \Lambda_{4d}$, the Higgs mechanism kicks in, breaks the gauge group and freezes the gauge coupling at a small value. In this regime we can trust the description of vortices given above, and use this as a valid starting point for semi-classical quantization of the string. The two-dimensional  $\CP^{N-1}$ sigma-model on the vortex worldsheet should now be quantized and, at scales below $ev$, we may talk about the running of the sigma-model coupling,  $1/r$. This will hit strong coupling at a scale $\Lambda_{2d}$. In the models that we consider below, we will have $\Lambda_{2d} \leq \Lambda_{4d}$.

\para
In Sections 3 and 4, we will see that it is fruitful to compare physics in the different regimes, $ev \ll \Lambda_{4d}$ and $ev \gg \Lambda_{4d}$. In particular, thanks to supersymmetry, certain quantities will be independent of the ratio of these scales. This is ultimately what is responsible for allowing us to compute certain 4d quantities through a study of the vortex worldsheet.

\subsubsection*{Generalizations}

The construction above describes the simplest theory
which gives rise to the $\CP^{N-1}$ sigma-model on the vortex worldsheet. We can now
start to play certain games. We can change the four-dimensional theory and ask how the vortex
worldsheet reacts. For example, one could add extra fields in four dimensions. It has long been known that 4d fermions, if coupled appropriately, donate massless 2d fermions to the vortex string \cite{jr,erick}. It is also true, that in certain circumstances bosonic fields may also induce extra zero mdoes \cite{semi2}.

\para
We could also ask more subtle questions. For example, suppose we add Yukawa couplings, or scalar potentials to the 4d theory. How do they affect the vortex dynamics? What happens if we turn on expectation values for further scalars in 4d? By understanding how each of these deformations affects the vortex worldsheet, we start to build a map from a large class of four-dimensional theories to a large class of two-dimensional sigma models.

\para
In Sections 3 and 4, we will describe in detail how many of the deformations mentioned above affect the dynamics of vortex strings in theories with ${\cal N}=2$ and ${\cal N}=1$ supersymmetry respectively. For now we will study a particular deformation of the theory that highlights further classical soliton solutions.

\subsection{Confined Monopoles}

We will consider the deformation that arises from assigning complex masses $m_i$ to the fundamental fields $q_i$. At the same time, we  introduce a further 4d scalar field which we denote as $a$: it is a complex scalar, transforming in the adjoint of the $U(N_c)$ gauge group. The resulting 4d Lagrangian is
\be
{\cal L}&=& \frac{1}{2e^2}{\rm Tr}(- F_{\mu\nu}F^{\mu\nu}+2|{\cal D}_\mu a|^2)
+\sum_{i=1}^{N_f}|{\cal D}_\mu q_i|^2 \nn \\
&&  - \frac{e^2}{2}{\rm Tr}(\ \sum_{i=1}^{N_f}q_iq_i^\dagger-v^2)^2 -
\sum_{i=1}^{N_f}q_i^\dagger\{a-m_i,a^\dagger-m^\dagger_i\}^2q_i\ .
\label{theorytwo}\ee
The mixing between the masses and the coupling to the adjoint scalar $a$ in the last term is of the type that arises in ${\cal N}=2$ supersymmetric theories. For now, though, we need only this bosonic action. We work, once again, with $N_f=N_c\equiv N$. The theory still has a unique vacuum, now given by
\be
q^a_{\ i}=v\delta^a_{\ i}\ \ \ \ ,\ \ \ \ a={\rm diag}(m_1,\ldots,m_N)\ .\label{a}\ee
We want to figure out how the vortex string reacts to the presence of non-zero masses $m_i$. Before doing any calculations, we can anticipate the effect. The $\CP^{N-1}$ orientational modes of the vortex arose from sweeping out with the $SU(N)_{\rm diag}$ symmetry preserved in the vacuum \eqn{coset}. But when $m_i\neq 0$, the $SU(N_f)$ flavor symmetry is explicitly broken. We therefore expect that, for generic masses, the orientational modes of the vortex will be lifted. Lets see how this works.

\para
In the presence of masses $m_i$, the surviving vortex solutions are those whose energy is not increased. For this to be the case, we require that the final term in \eqn{theorytwo} vanishes when evaluated on the vortex solution, with $a$ in its vacuum state. This, in turn, requires that $(a-m_i)q_i=0$. For generic $m_i$, there are precisely $N$ such solutions which arise when the Abelian vortex is embedded diagonally within the gauge group,
\be
B={\rm diag}(0,\ldots,B^\star,\ldots,0)\ \ \ ,\ \ \ q={\rm diag}(v,\ldots,q^\star,\ldots,v)\ . \label{bstar}\ee
In this way, the single topological quantum number $k$ gets split into $N_c$ distinct low-energy quantum numbers, distinguished by the diagonal element of $B$ that carries the flux and by the flavour $q_i$ that winds around the string.

\subsubsection*{The View from the String}

We can describe this lifting of orientational modes from the perspective of the vortex worldsheet \cite{memono,sy,vstring}.
The masses $m_i$ can
be thought of as inducing a potential over the ${\bf CP}^{N-1}$ moduli space whose zeroes correspond to the surviving solutions. This potential can be shown to take a simple
geometrical form, being proportional to the length of a particular Killing
vector on ${\bf CP}^{N-1}$ \cite{vstring}. For a single vortex string, it is a simple matter to describe the outcome of this procedure using the homogeneous coordinates introduced in \eqn{cpn}. We introduce an auxiliary complex field, $\sigma$, on the vortex worldsheet. The vortex dynamics is described by the Lagrangian,
\be {\cal L}_{\rm vortex} = \ft12T|\partial_\alpha Z|^2 + |{\cal D}_\alpha\varphi_i|^2 + \lambda\left(|\varphi_i|^2-r\right) - \sum_{i=1}^N|\sigma-m_i|^2|\varphi_i|^2\ ,\label{v2}\ee
where the masses $m_i$ on the worldsheet are identified with the masses $m_i$ introduced in four dimensions. We see that the orientational modes are indeed lifted by the masses, and the vortex theory has $N$ isolated vacuum states given by,
\be |\varphi_j|^2=r\delta_{ij}\ \ \ ,\ \ \ \sigma=m_i\ \ \ \ i=1,\ldots,N\ .\label{csig}\ee
Notice that, although we have $N$ different types of vortex \eqn{bstar}, these are all described within a single worldsheet theory where they appear as different ground states. This fact allows us to find something new.

\subsubsection*{The Kink and the Monopole}

Since the string has multiple ground states, we have a new object in the theory: a kink, interpolating between different worldsheet vacua at $x^3\rightarrow -\infty$ and $x^3\rightarrow +\infty$ \cite{memono}. From the perspective of the worldsheet theory, this object is BPS, obeying the following first order equations:
\be {\cal D}_3\varphi_i=(\sigma-m_i)\varphi_i\ \ \ ,\ \ \ \sigma= \frac{1}{r}\sum_{i=1}^{N} m_i|\varphi_i|^2\ . \ee
One can think of this kink as a bead threaded on the vortex string, able to move freely back and forth. But what is the interpretation of the bead? Let us first look at the quantum numbers of the kink, specializing to $U(2)$ gauge theory for simplicity. The kink interpolates between two strings, each carrying magnetic flux in a different $U(1)$ subgroup, as shown in the Figure 3. The kink must soak up magnetic flux $B={\rm diag}(0,1)$ and spit out magnetic flux $B={\rm diag}(1,0)$. In other words, it is a source for magnetic flux $B={\rm diag}(1,-1)$. This is the same magnetic charge as a 't Hooft-Polyakov monopole in  $SU(2) \subset U(2)$.

\para
We can compute the  mass of the bead. For a kink interpolating between the $\sigma=m_i$ and $\sigma=m_j$ string vacuum, the mass is given by,
\be M_{\rm kink} = r\Delta \sigma = r|m_i-m_j|\ .
\ee
\begin{figure}[htb]
\begin{center}
\includegraphics*[width=9cm]{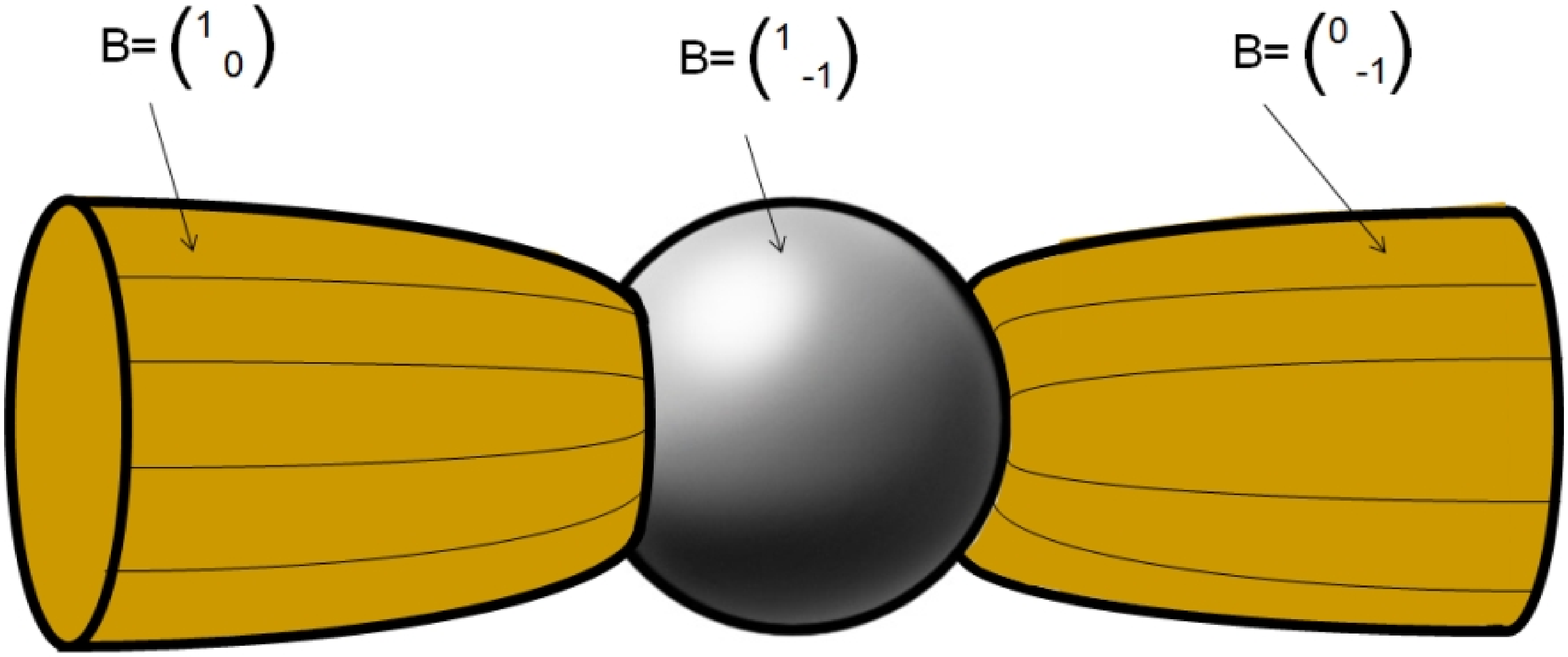}
\caption{%
A cartoon of the magnetic monopole emitting two flux tubes. The size of the monopole is $L_{\rm mono}\sim 1/\Delta m$. The width of the flux tube is $L_{\rm vort} \sim 1/ev$.}
\label{figtwo}
\end{center}
\end{figure}
We can translate this into 4d quantities using \eqn{r} and \eqn{a}, to describe the mass in terms of the expectation values of the adjoint scalar field $a$,
\be M_{\rm kink} = \frac{4\pi}{e^2} |a_i-a_j| = M_{\rm mono}\ ,\label{km}\ee
which we recognize as the mass $M_{\rm mono}$ of a BPS 't Hooft-Polyakov monopole in $SU(N)$ gauge theory.

\para
The bead on the string looks and smells like a magnetic monopole. In fact, the physics behind this is clear. The usual 't Hooft-Polyakov monopole lives in the Coulomb phase, with the gauge group  broken to  $U(1)$ factors. The magnetic flux escapes radially to infinity and is captured by the integral over the ${\bf S}^2_\infty$ boundary.  In our model, this occurs when $v^2=0$, and $\langle a \rangle \neq 0$. In contrast, when $v^2 \neq 0$, the theory lies in the Higgs phase and the gauge bosons are massive. The theory is a non-Abelian superconductor, exhibiting the Meissner effect. The flux of the magnetic monopole cannot propagate in the vacuum, and so instead leaves in two collimated flux tubes. These flux tubes are the vortex strings. The monopole is confined.

\para
While the existence of the magnetic monopole is guaranteed by the topology and
symmetry breaking of the theory, there is something rather special about the mass formula \eqn{km}. As we turn on $v^2$, the mass of the monopole jumps from a finite number, $M_{\rm mono}$, to infinity, reflecting its confinement and the presence of the two semi-infinite flux tubes. However, even in the confined phase, we can still assign a finite mass to the monopole, by viewing it as a finite energy excitation above an infinite vortex string. Rather surprisingly, equation \eqn{km} tells us that,  in our model, this finite excitation is independent of the string tension $v^2$ and, in particular, remains constant as $v^2\rightarrow 0$. Ultimately, this can be understood as a non-renormalization theorem arising from an underlying ${\cal N}=2$ supersymmetry \cite{sy,vstring}.

\subsubsection*{Back to the View From Spacetime}

Having discovered the confined monopoles from the perspective of the vortex string, we now return to 4d spacetime and ask what the configurations look like. The configurations are BPS and one can derive first order Bogomolnyi equations \cite{memono},  which can be shown to imply solutions to the equations of motion arising from \eqn{theorytwo}. They are
\be
B_1={\cal D}_1a\ \ \ ,\ \ \ B_2={\cal D}_2a\ \ \ ,\ \ \
B_3={\cal D}_3a+e^2(\sum_{i=1}^Nq_iq_i^\dagger -v^2)\nn\\
{\cal D}_1q_i=i{\cal D}_2 q_i\ \ \ ,\ \ \
{\cal D}_3 q_i=-(a-m_i)q_i\hspace{2.5cm}
\label{boggy}\ee
These equations combine the vortex equations (to which they reduce when $B_1=B_2={\cal D}_3q_i=0$) with the BPS monopole equations (when $v^2=0$). No explicit solutions to these equations are known. Indeed, the system of differential equations is over determined. Nonetheless, they are expected to admit solutions which are monopoles threaded on vortex flux-tubes, as described above.

\para
The size of the flux-tube is set by $L_{\rm vortex} = 1/ev$ while the size of the monopole is set by $L_{\rm mono}=1/\Delta m$. The worldsheet description of the confined monopole is valid in the regime $L_{\rm vortex} \ll L_{\rm mono}$. In contrast, when  $L_{\rm vortex} \gg L_{\rm mono}$, we should look to the full 4d configuration. The flux from the magnetic monopole spreads out radially until it reaches the penetration depth $L_{\rm vortex}$ of the non-Abelian superconductor, at which point it forms collimated flux tubes \cite{memono,sy}. Far from the monopole, these flux tubes are the vortex strings.

\subsection{Other BPS Solitons}

So far we have discussed 4d theories with $N_f=N_c$. When $N_f > N_c$, the 4d theory has multiple vacuum states resulting in yet another class of 4d solitons: domain walls. The first order equations describing these objects are again given by \eqn{boggy}, this time by setting $B={\cal D}_1q={\cal D}_2q=0$.

\para
There is fairly compelling evidence that the full equations \eqn{boggy} contain solutions which reflect an intricate set of interactions between the domain walls, the vortex strings and the confined monopoles. The vortices can end on domain walls, providing a field theoretic realization of D-branes. As we have seen, monopoles are threaded on vortices. If these vortices end on walls, there are rules governing which monopole can pass through which walls. I will not describe these results in detail here but detailed reviews of these solitons can be found in \cite{booj,tasi,mmatrix,syreview}. Since a picture paints a thousand words, I will restrict myself here to showing a beautiful depiction of field theoretic D-branes taken from \cite{allquarter} in which the equations \eqn{boggy} are solved numerically (in the $e^2\rightarrow \infty$ limit).
\begin{figure}[htb]
\begin{center}
\epsfxsize=3.2in\leavevmode\epsfbox{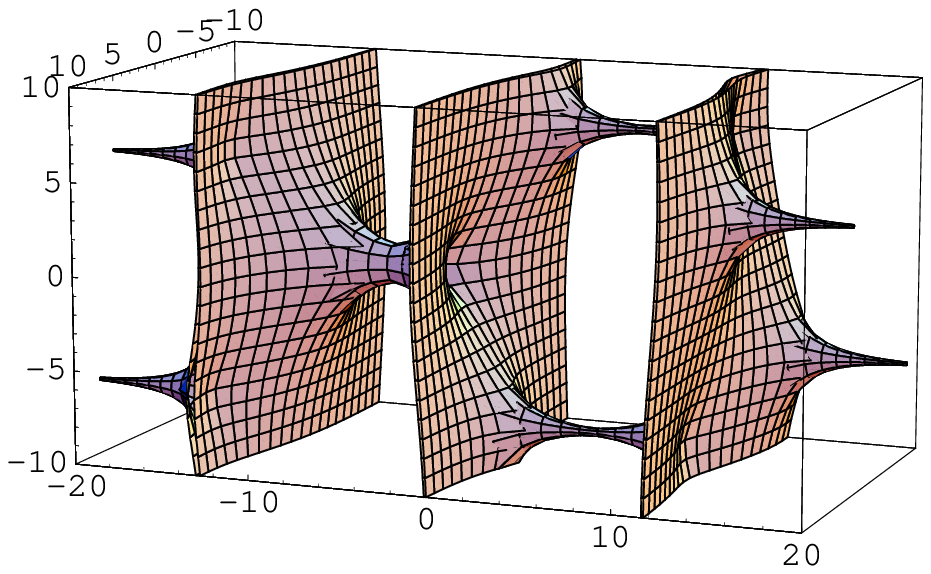}
\end{center}
\caption{Plot of a field theoretic D-brane configuration
\cite{allquarter}.}
\end{figure}

\subsection{Applications}

Sections 3 and 4 of this review will focus on the quantum dynamics of vortex strings in supersymmetric gauge theories. However, in recent years a large literature has developed studying various properties and applications of non-Abelian vortex strings which are unrelated to supersymmetry. I summarize some of these results here.

\para
In the cosmological context, the internal modes of non-Abelian vortices affect the probability for reconnection of cosmic strings which, in turn, affects the density of string networks seen in the sky. The reconnection of non-Abelian cosmic strings was discussed in \cite{2vortex,2v2,2v3}. Related work, studying the forces between vortices with different orientation when the potential deviates from its critical value can be found in \cite{nonsusy}. Finally, the interaction of non-Abelian vortices with axions was studied in \cite{axion}.

\para
In the context of QCD, non-Abelian {\it chromo-magnetic} strings have been suggested to be of importance at high densities \cite{nitta,hot}. At high temperatures, non-Abelian strings share some key characteristic properties with the ${\bf Z}_N$ magnetic strings seen in lattice simulations \cite{gz}, which are argued to form part of the Yang-Mills plasma \cite{cz}. Indeed, the strings seen in an Abelian projection of the lattice data also have beads of monopoles threaded on them in a manner reminscent of the discussion in Section 2.1 above \cite{cz2}. However, it should be noted that the monopoles in the lattice are junctions where $N$ strings meet, while the BPS monopoles described here are always junctions for two strings. Further aspects of ${\bf Z}_N$ strings which carry orientational modes were discussed in \cite{other}.

\para
Finally, the non-Abelian vortices have also been suggested as a good model of {\it chromo-electric} strings, in the spirit of dual-confinement first proposed by Nambu, 't Hooft and Mandelstam \cite{th,mand}. From this viewpoint, the non-Abelian monopoles provide a model of quarks and the orientational modes of the vortex carry genuine non-Abelian flux. If this were true, the effects of the orientational modes would show up in lattice simulations of the QCD string --- for example, through the L\"uscher term \cite{luscher}. Note that although the monopoles described above are BPS and emit two flux tubes in opposite directions, by embedding the $U(N)$ theory in a larger rank gauge group (for example, $SU(N+1)\rightarrow U(N)$), one can show that there exist monopoles on which a single string can end \cite{nonmon,abe}. A dual model of color confinement in the context of ${\cal N}=1$ supersymmetric theories was presented in \cite{etoal}. Nice reviews of some of the ideas concerning non-Abelian monopoles and their dual realization can be found in \cite{konrev,kon1}.

\section{Vortex Strings and Seiberg-Witten Theory}

The low-energy physics of four dimensional $SU(2)$ gauge theories with ${\cal N}=2$ supersymmetry was solved some years ago by Seiberg and Witten \cite{sw,sw2}, with generalizations to $SU(N_c)$ theories given in \cite{ho,apshap}. The theory does not exhibit confinement, but has a slew of other interesting properties including the presence of massless monopoles and novel interacting superconformal field theories. The Seiberg-Witten solution describes the interactions of the massless degrees of freedom (it is, in some sense, analogous to computing $f_\pi$ in QCD from first principles) and also allows us to compute the exact mass spectrum of BPS states in the theory. I will not review the Seiberg-Witten solution here --- good reviews can be found in \cite{swrev}. Rather, in this section, we will see how the Seiberg-Witten results can be re-derived through a study of the dynamics of the vortex string.

\para
The Lagrangian \eqn{theorytwo} is a subset of the Lagrangian for ${\cal N}=2$ super QCD. (It is missing $N_f$ anti-fundamental scalars, $\tilde{q}_i$, as well as the fermions). The Seiberg-Witten solution holds when $v^2=0$ and for $SU(N_c)$ gauge group rather than $U(N_c)$. However, since the $U(1)\subset U(N_c)$ is infra-red free, this last point makes little difference.
The Seiberg-Witten solution is typically presented as a function over the Coulomb branch of vacua, parameterized by the adjoint scalar expectation value $\langle a \rangle$, with fixed masses $m_i$. Here we will instead focus attention on a particular point of the Coulomb branch. In the theory with $N_f=N_c$, this point is defined classically as
\be \langle a \rangle= {\rm diag} (m_1,\ldots,m_N)\ .\label{cvac}\ee
At this point, there are $N$ massless quarks. This means that expectation value $\langle a \rangle$ is left undisturbed as we turn on $v^2$ and push the theory into the Higgs vacuum \eqn{a} where these $N$ massless scalar quarks condense. For this reason, the point \eqn{cvac} on the Coulomb branch is sometimes called the ``root of the baryonic Higgs branch". While the vortex string exists in the Higgs phase, as we explained in the introduction, the core of the vortex string will know about 4d physics at the point \eqn{cvac} on the Coulomb branch.  Although we will not be able to explore the full Coulomb branch through the eyes of the vortex string, by varying the masses $m_i$ we will find that we can reproduce many of the interesting phenomena that occur in Seiberg-Witten theory.

\para
The point \eqn{cvac} is the classical root of the baryonic Higgs branch. However, in the quantum theory, this point gets shifted \cite{aps}. One can show that the point on the Coulomb branch where $N$ quarks become massless, is actually given by the expectation values $\langle a\rangle = {\rm diag}(a_1,\ldots,a_{N_c})$ which satisfy the following equation\footnote{For aficionados, the Seiberg-Witten curve at this point degenerates, reflecting the existence of $N$ massless quarks. It is given by $y^2=\left(\prod_{i=1}^{N}(x-m_i)-\Lambda^N\right)^2$} for all values of $x$:
\be \prod_{b=1}^{N}(x-a_b)=\prod_{i=1}^{N}(x-m_i)+\Lambda_{4d}^{N}\ .\label{qvac}\ee
Here the strong coupling scale $\Lambda_{4d}$ is given in terms of the 4d gauge coupling $e^2$, defined at the RG subtraction point $\mu$. The beta-function is exact at one-loop in this theory, and gives
\be \Lambda^{N}_{4d}=\mu^{N}\exp\left(-\frac{4\pi^2}{e^2(\mu)}\right)\ee
In the weak coupling regime, when $\Delta m_i \gg \Lambda_{4d}$, the quantum vacuum \eqn{qvac} coincides
with the classical vacuum \eqn{cvac}. In contrast, at strong coupling when $\Delta m_i\ll \Lambda_{4d}$, the expectation values huddle around $\Lambda_{4d}$.


\subsection{The ${\cal N}=(2,2)$ Vortex Worldsheet}\label{qworld}

Let us now turn to the vortex worldsheet. As we outlined in the introduction, we deform the theory by turning on $v^2$ until $ev\gg \Lambda_{4d}$, so that the theory is frozen at weak coupling.

\para
The bosonic part of the vortex dynamics is exactly the same as that described in equation \eqn{v2} of Section 2.1. However, it is now augmented by fermions, originating from the zero modes associated to the fermions in four-dimensions \cite{jr}. The resulting worldsheet dynamics has ${\cal N}=(2,2)$ supersymmetry in two dimensions; this is four supercharges, and is to be expected since the vortex string is $\ft12$-BPS in the four-dimensional ${\cal N}=2$ theory.

\para
There are two Goldstino modes,  $\chi_\pm$, arising from the supercharges that are broken in the presence of the vortex. These sit, together with the translational collective coordinate $Z$, in a ${\cal N}=(2,2)$ chiral multiplet. At lowest order in the derivative expansion, all these fields are free. More interesting are the internal modes and their fermionic partners. We introduce $N$ left-moving Weyl fermions $\xi_{-i}$, and $N$ right-moving Weyl fermions $\xi_{+i}$ on the the worldsheet. Each of these has charge $+1$ under the auxiliary worldsheet gauge field $u_\alpha$. The form of the worldsheet Lagrangian is fixed by ${\cal N}=(2,2)$ supersymmetry. It is
\be {\cal L}_{\rm vortex} &=& |{\cal D}_\alpha\varphi_i|^2 +
2i\left(\bar{\xi}_{+i}{\cal D}_-\xi_{+i} + \bar{\xi}_{-i}{\cal D}_+\xi_{-i}\right) \nn\\
&& - |\sigma-m_i|^2|\varphi_i|^2 + (\bar{\xi}_{-i}(\sigma-m_i)\xi_{+i}+ {\rm h.c.})
\nn\\ && + \lambda\left(|\varphi_i|^2-r\right)  + (\bar{\zeta}_-\bar{\xi}_{+i}\varphi_i+\bar{\zeta}_+\bar{\xi}\varphi_i+{\rm h.c.}) \label{v22}
\ee
where all summations over $i=1,\ldots,N$ have been left implicit. The derivatives acting on fermions are ${\cal D}_\pm= {\cal D}_0\mp{\cal D}_3$.

\para
In the final line of \eqn{v22}, we have introduced two Grassmann Lagrange multipliers, $\zeta_+$ and $\zeta_-$. These are the superpartners of auxiliary bosonic fields $\lambda$ and $\sigma$. Their role is to impose the constraints that the fermi zero modes live in the tangent bundle of $\CP^{N-1}$ or, simply,
\be \sum_{i=1}^N \bar{\xi}_{+i}\varphi_i= \sum_{i=1}^N\bar{\xi}_{-i}\varphi_i=0
\ee

\subsubsection*{The Quantum Worldsheet}

We now describe the quantum theory on the vortex worldsheet. The first hint that there is something magical going can be seen by studying the beta-functions. Recall from \eqn{r} that the 2d sigma-model coupling and 4d gauge coupling are related classically,
\be r=\frac{4\pi}{e^2} \label{re}\ee
Importantly, in theories with ${\cal N}=2$ supersymmetry, this relation is preserved under RG flow. Both 2d and 4d beta-functions are one-loop exact. If we first set the masses to zero, $m_i=0$, then the running of the coupling is given by
\be r(\mu) = r(\mu_0)-\frac{N}{2\pi}\log\left(\frac{\mu_0}{\mu}\right)\label{rrun}\ee
This ensures that the running of the 2d coupling tracks the would-be running of the 4d coupling (``would-be" because we froze the coupling $e^2$ at the Higgs scale). This has the consequence that the 2d and 4d strong coupling scales agree: $\Lambda_{2d}=\Lambda_{4d}$.

\para
The running coupling \eqn{rrun} holds when the masses vanish: $m_i=0$. Turning on the masses breaks the non-Abelian global symmetry of the ${\bf CP}^{N-1}$ sigma-model: $SU(N)\rightarrow U(1)^{N-1}$. This halts the running of the 2d coupling $r$. This is entirely analogous to the situation in 4d where the expectation value $\langle a \rangle \sim {\rm diag}(m_1,\ldots,m_N)$ breaks the gauge group $U(N)\rightarrow U(1)^N$, halting the running of the 4d coupling $e^2$. The upshot of this is that for $\Delta m_i \gg \Lambda_{2d} = \Lambda_{4d}$, both the 2d worldsheet theory and the 4d gauge theory are weakly coupled.

\para
We now turn to the mass spectrum of the worldsheet theory. The spectrum of the bosonic and supersymmetric  $\CP^{N-1}$ sigma-model was solved many years ago in \cite{dadda,beaut} and \cite{phases}. The solution of the model with masses $m_i$ was solved in \cite{nick}. The key idea in all of these papers is to figure out what fields to focus on. Naively, you might think that you're interested in the dynamical fields $\varphi^i$ and $\xi_{\pm i}$. But you're not! You're really interested in the constraint and auxiliary fields. These are the gauge field $u_\alpha$, the Lagrange multipliers $\lambda$ and $\zeta_\pm$ and the auxiliary complex scalar $\sigma$. They live in the same supermultiplet which,  in two dimensions, forms a  twisted chiral superfield, $\Sigma$
\be \Sigma = \sigma + \theta^+\bar{\zeta}_++\bar{\theta}^-\zeta_-+\theta^+\bar{\theta}^-(\lambda-iG_{03})\ee
where $G_{03}=\partial_0u_3-\partial_3u_0$ is the auxiliary field strength.

\para
To compute the mass spectrum of the theory, we integrate out $\varphi_i$ and $\xi_{\pm i}$  to derive an effective action for the superfield $\Sigma$. The leading contributions to this effective action are the kinetic term, and potential term which, in superfield language, are written in terms of a K\"ahler potential $K(\Sigma,\Sigma^\dagger)$ and a holomorphic (twisted) superpotential ${\cal W}(\Sigma)$,
\be {\cal L}_{eff} &=& \int d^4\theta K(\Sigma,\Sigma^\dagger) + \int d\bar{\theta}^+ d\theta^- {\cal W}(\Sigma) + {\rm h.c.} \nn\\ &=& \frac{\partial^2 K}{\partial \sigma\partial\sigma^\dagger} |\partial_\alpha\sigma|^2 - \left(\frac{\partial^2 K}{\partial \sigma\partial\sigma^\dagger} \right)^{-1}\left|\frac{\partial {\cal W}}{\partial \sigma}\right|^2 + {\rm fermions}\ee
The K\"ahler potential is unknown (it can be computed at weak coupling when $\Delta m \gg \Lambda$). However, the superpotential is fixed exactly by holomorphy and various symmetries. It includes a tree-level piece, and one-loop piece, and is given by \cite{phases,hh},
\be {\cal W}(\sigma) = -r\sigma -\frac{1}{2\pi}\sum_{i=1}^N(\sigma-m_i)\left[\log\left(\frac{\sigma-m_i}{\mu}\right)-1 \right] \label{310}\ee
The vacua of the vortex string sit at the critical points of ${\cal W}$. We find that the $N$ vacua of the classical theory survive to the quantum theory (indeed, the Witten index tells us that they must), but they are shifted from their classical values:
\be \frac{\partial {\cal W}}{\partial \sigma}=0\ \ \ \Rightarrow \ \ \ \prod_{i=1}^N (\sigma-m_i)=\Lambda_{2d}^N\label{crut}\ee
At weak coupling, $\Delta m_i \gg \Lambda_{2d}$, the vacua are close to their classical values \eqn{csig}. However, when the sigma-model is strongly coupled, $\Delta m_i \ll \Lambda_{2d}$, the worldsheet vacua gather around the strong coupling scale $\Lambda_{2d}$.

\para
How do we interpret these vacua at strong coupling? To see this, let us look at the equation of motion for $\sigma$ arising from \eqn{v22},
\be ir\sigma = \sum_{i=1}^N (\bar{\xi}_{+i}\xi_{-i} + m_i|\varphi_i|^2)\label{whatsigma}\ee
In the absence of masses, $\sigma$ controls the fermi condensate of the theory. The $N$ vacuum states are then the $N$ different values for the condensate of fermi zero modes on the vortex string.

\subsubsection*{Quantum Kinks}

Since the quantum vortex string still has $N$ isolated ground states, we may look at the kinks on the string. As we argued in Section 2, these are interpreted as magnetic monopoles, confined to live as beads threaded on the vortex. Even though we do not know the value of the K\"ahler potential $K(\sigma)$, we still have enough information to compute the exact mass of these kinks in the quantum theory. To see this, we simply complete the square in the energy functional for the string, \`a la Bogomolnyi. We concentrate on static field configurations. Writing $f(\sigma,\sigma^\dagger)=\partial^2K/\partial\sigma\partial\sigma^\dagger$,  the mass of the kink is given by,
\be E &=& \int dx^3 \ f|\sigma'|^2 + f^{-1}\left|\frac{\partial {\cal W}}{\partial \sigma}\right|^2 \nn\\ &=& \int dx^3\ f \left|\frac{\partial\sigma}{\partial x^3} - \alpha f^{-1}\left(\frac{\partial {\cal W}}{\partial \sigma}\right)^\dagger\right|^2 +\alpha^\star \frac{\partial \sigma}{\partial x^3} \frac{\partial{\cal W}}{\partial \sigma} + \alpha \frac{\partial \sigma^\dagger}{\partial x^3} \frac{\partial{\cal W}^\dagger}{\partial \sigma^\dagger}
\nn\\ & \geq & 2 {\rm Re}\left[\alpha^\star \left.{\cal W}\right|_{x^3=+\infty} -
\alpha^\star \left.{\cal W}\right|_{x^3=-\infty}\right]\ ,
\ee
for some phase $\alpha$. We can maximize the inequality by choosing $\alpha = \Delta {\cal W}/|\Delta {\cal W}|$. This maximal inequality is then saturated by BPS kinks, whose mass is given by,
\be M_{\rm kink} = 2|\Delta {\cal W}|\ .\ee
In this way, the mass of the kink on the vortex string can be computed analytically for all values of the masses $m_i$ using the exact superpotential \eqn{310}.

\para
We can compare this calculation to the exact quantum mass of the 4d BPS monopole computed using the  Seiberg-Witten solution \cite{sw,sw2}. The relevant calculation is rather different to the one described above: one must integrate a particular one-form over a cycle in the Seiberg-Witten curve\footnote{Those familiar with the Seiberg-Witten solution will note that the Seiberg-Witten curve has a natural expression in terms of the 2d superpotential: $y^2=(\partial {\cal W}/\partial x)^2$. This is the key mathematical statement underlying the physical agreement between the two theories.}.  Remarkably, the result agrees with the computation of the kink mass described above, so that even in the quantum theory we have the result \cite{nick,sy,vstring}
\be M_{\rm kink} = M_{\rm mono}\ .
\label{mkmm}\ee

\subsubsection*{Yang-Mills Instantons vs Worldsheet Instantons}

The derivation above gives the exact quantum mass spectrum of the kinks. However, it is deceptively simple and it is illustrative to recast the answer in a more explicit form. Consider, for example, the $U(2)$ gauge theory and set $m_1=-m_2=m$. The theory is weakly coupled when $|2m|\gg\Lambda$. In this regime, we may expand the
soliton mass as
\be M_{\rm kink} = M_{\rm mono} &=& \frac{2}{\pi}\left|\sqrt{\Lambda^2+m^2} + \frac{m}{2} \log\left(\frac{\sqrt{\Lambda^2+m^2}-m}{\sqrt{\Lambda^2+m^2}-m}\right)\right|\nn\\ &=& \frac{2|\,m|}{\pi}\left|\left(1+\log\left(\frac{\Lambda}{2m}\right)+\sum_{n=1}^\infty c_n \left(\frac{\Lambda}{2m}\right)^{2n}\right)\right|\ .\ee
The semi-classical interpretation of this expansion is clear. The first two terms describe the classical and one-loop contributions to the soliton mass, now written in terms of the RG invariant scale $\Lambda$ rather than $r$. Although supersymmetry forbids higher loop corrections, there are an infinite series of worldsheet instanton contributions, with coefficients $c_n$ which can be easily computed from the Taylor expansion of the exact result. This is an expansion in $\Lambda$ or, alternatively,
\be \exp(-S_{\rm inst})= \exp(-2\pi r) = \exp\left(-\frac{8\pi^2}{e^2}\right)\ee
In the two-dimensional theory, these instanton contributions to the mass $M_{\rm kink}$ are due to worldsheet instantons of the ${\bf CP}^{N-1}$ sigma-model; in the four-dimensional theory, the contributions to $M_{\rm mono}$ are from $U(N)$ Yang-Mills instantons. And yet, term by term, the two series agree.

\para
This agreement between the 2d and 4d instanton expansions is due, in part, to the fact that the two instantons represent the same object. The bulk of the 4d theory lies in the Higgs phase. Here the instanton shrinks to zero size and no non-singular solution exists. However, the Yang-Mills instanton may shelter from the Higgs expectation value by nestling inside the core of the vortex string, where it appears in the guise of a  worldsheet instanton\footnote{This is particularly clear from the brane picture described in \cite{vib,vstring}, where both objects arise as Euclidean D0-branes lying in the D4 world-volume.}.
This is analogous to
the fact that, as we have seen above, a Yang-Mills monopole appears as a kink on the vortex string. Indeed, just as there exist first order Bogomolnyi equations \eqn{boggy} describing the confined monopole, so there also exist first order equations describing Yang-Mills instantons nestled inside the vortex string. We set the masses $m_i=0$ to zero and work in four-dimensional Euclidean space. Defining a complex structure on ${\bf R}^4$ to be given by $z=x^1+ix^2$ and $\omega=x^4+ix^3$, the relevant equations are
\be
F_{12}-F_{34}= e^2\sum_iq_iq_i^\dagger - e^2v^2 \ \ ,\ \  F_{14}=F_{23}\ \ , \ \ F_{13}=F_{24} \ \ ,\ \  {\cal D}_zq_i=0\ \ , \ \
{\cal D}_{\bar{w}}q_i=0
\nn\ee
These equations are a mixture of the self-dual Yang-Mills equations and the non-Abelian vortex equations. In the physics literature, they were first derived in \cite{vstring} and studied further in \cite{ihiggs}. However, they were earlier introduced in the mathematics literature by Taubes in order to prove the equivalence of Gromov invariants and Seiberg-Witten invariants for symplectic four-manifolds \cite{ct}. The quantum equivalence between 2d and 4d theories described above appears to be the physical incarnation of Taubes' result.

\para
While the exact results above demonstrate agreement between the instanton expansions in 2d and 4d, one can ask how this might arise from a first-principles microscopic semi-classical calculation. This remains an open question. However, a clue certainly lies in the observation that the moduli space of $k$ worldsheet instantons in ${\bf CP}^{N-1}$ --- which we denote as $\widehat{\cal V}_{k,N}$ --- is a complex submanifold of ${\cal I}_{k,N}$, the moduli space of $k$ non-commutative Yang-Mills instantons in $U(N)$ \cite{vib}.
As shown explicitly by Nekrasov \cite{nek}, localization
theorems hold when performing the integrals over ${\cal I}_{k,N}$ in ${\cal N}=2$ SQCD
and the final answer contains contributions from only a finite number
of points in ${\cal I}_{k,N}$. It is simple to check that all of these points lie on $\widehat{\cal V}_{k,N}\subset {\cal I}_{k,N}$. It would be interesting to put these
observations on a firmer footing.

\subsubsection*{W-Bosons, Dyons, and Curves of Marginal Stability}

Agreement between the quantum spectra of the 2d and 4d theories stretches beyond equation \eqn{mkmm}. Other results include:

\para
$\bullet$ The elementary internal excitations of the string can be identified
with W-bosons of the 4d theory. When in the bulk, away from the string, these
W-bosons are non-BPS. But they can reduce their mass by taking refuge in the
core of the vortex whereupon they regain their BPS status \cite{sy,vstring}.

\para
$\bullet$ The 4d theory
contains dyons. The 2d theory also
contains dyonic kink objects, first described classically in \cite{at}. The quantum corrected masses of dyons in the two theories are identical \cite{nick}. Moreover, for $N_c\geq 3$, both the topological and Noether charges become ($N_c-1$)-vectors. Both 4d theories and 2d theories have an interesting spectrum of states where the topological charge vector does not lie parallel to the Noether charge vector and these dyons have an interesting structure of decay modes as the masses $m_i$ are varied and one crosses curves of marginal stability. Recent work has studied this decay in the 4d theory and its relationship to the  Kontsevich-Soibelman formula \cite{moore}. It would be interesting to match this to the decay of dyons in the 2d worldsheet theory.

\para
$\bullet$ The theta angle in 4d can be shown to induce a 2d theta angle on the vortex worldsheet \cite{vstring,gsy}. In both 2d and 4d, the coupling constants are naturally complexified,
\be t = r+i\theta = \frac{4\pi}{e^2}+\theta\ee
Both theories then manifest the Witten effect, in which the theta angle
induces a Noether charge on the soliton, shifting its mass \cite{witteff,nick}.

\para
$\bullet$ We have here described the theory with $N_f=N_c$. For $N_f>N_C$, the nature of the vortex strings is qualitatively different: they pick up (logarithmically divergent) scaling modes, and are sometimes referred to as ``semi-local" strings \cite{semi2}. They were studied in non-Abelian gauge theories in \cite{vib,shifsemi}. The story described above can be repeated, and the worldsheet spectrum once again coincides with the spectrum of the 4d theory in which it's embedded
\cite{dht,vstring}.

\subsection{Superconformal Worldsheets}

One of the most striking features to arise from Seiberg and Witten's analysis of ${\cal N}=2$ gauge theories is the existence of points on the Coulomb branch where magnetic monopoles become massless. How is this seen from the perspective of the vortex string?

\para
Because we are sitting on Coulomb branch at \eqn{qvac}, where massless quarks already exist, if we can tune the $m_i$ so that monopoles also become massless then the 4d theory would have massless degrees of freedom with both magnetic charge and electric charge. This results in an interacting superconformal field theory (SCFT), known as Argryes-Douglas point \cite{ad}.

\para
For the case of $N_f=N_c\equiv N$ , massless monopoles can be shown to  arise when the masses are tuned to the critical point,
\be m_k=-\exp\left(2\pi i k/N\right)\Lambda\ \ \ \ \ ,\ \ \ \ \ k=1,\ldots, N\label{critm}\ee
The 4d SCFT at this point fits into the ADE classification of \cite{ehiy,eh} --- it is the $A_{2N-1}$ theory in the language of those papers.

\para
Since we have $M_{\rm mono}=M_{\rm kink}$, the masses of kinks on the vortex worldsheet must also vanish \cite{zwicky}. Indeed, this is simple to see. From \eqn{crut}, the $N$ isolated worldsheet vacua lie at $\prod(\sigma-m_i)=\Lambda^N$. At the critical point \eqn{critm}, all of these vacua coalesce at the origin $\sigma=0$. The question is: what is the physics of the worldsheet at this point? This is simple to answer \cite{scvs}. We expand the twisted superpotential \eqn{310} at the critical point for small $\sigma/\Lambda$ to find that the familiar logarithms vanish, and are replaced by a polynomial Landau-Ginzburg model,
\be {\cal W}(\sigma) = c_0\,\frac{\sigma^{N+1}}{\Lambda^N}+\ldots\ee
where $c_0$ is an overall constant coefficient, and $\ldots$ refer to irrelevant operators. The K\"ahler potential of the worldsheet theory is unknown at this strongly coupled point. However, this is unimportant because the K\"ahler potential is expected to adjust itself under RG flow so that the theory flows to an interacting ${\cal N}=(2,2)$ superconformal field theory  which is identified with the $A_{N-1}$ minimal model in two dimensions \cite{martinec,vafwar,lg}.

\para
It's interesting that the ${\bf CP}^{N-1}$ sigma-model --- the archetypal example of a theory with a demonstrable dynamical mass gap --- can be tuned to flow to an interacting fixed point. This is achieved by turning on a bare masses $m_i$ to precisely cancel the dynamically generated mass $\Lambda$.

\para
At the superconformal point, the equivalence of the 2d and 4d BPS mass spectra descends to an equivalence of the spectra of chiral primary operators. The dimensions of the relevant chiral perturbations of the 2d minimal model are given by,
\be D_j = \frac{j}{N+1}\ \ \ \ \ ,\ \ \ \ \ \ j=2,\ldots, N \label{vdim}\ee
The relevant deformations of the 4d theory fall into two distinct classes \cite{apsw}. The first class correspond to varying the mass parameters $m_i$, such that we stay within the class of vacua \eqn{qvac} where vortex strings exist. These deformations are seen on the vortex string and their dimension agrees with the 2d result \eqn{vdim}. The second class of operators involve a variation of the scalar vev $\langle a \rangle$, leaving the masses $m_i$ untouched. These take us away from the root of the baryonic Higgs branch \eqn{qvac} and are not seen directly on the vortex string. Nonetheless, as shown in \cite{apsw}, the dimensions of these operators are fixed by \eqn{vdim}, and are given by
\be \tilde{D}_{N-j+2} = 2 - D_j \label{infer}\ee
In this way, the 2d theory captures the information about the scaling dimensions of all 4d chiral primary operators \cite{scvs}: half of these are seen directly on the worldsheet; the remaining half are inferred through \eqn{infer}. The analogy between 2d and 4d spectra in the vicinity of a superconformal points was noted previously in \cite{shapv} and has been explored more recently in the context of vortex stings in \cite{ritz}.

\para
Note, however, that while the spectrum of chiral primary operators agree, other aspects of the 2d and 4d theories do not. For example, the central charge of the 2d SCFT is (ignoring the free, translational mode) $c_{2d} = 3 - 6/(N+1)$. Note that the number of degrees of freedom does not grow without bound as $N\rightarrow \infty$. Rather, the internal modes asymptote to a single free chiral multiplet which can be identified with $\Sigma$. In contrast, the central charges of the 4d theory was recently computed \cite{shap} and, as $N\rightarrow \infty$, scale as  $a_{4d}\sim c_{4d} \sim {\cal O}(N) $, reflecting the growing number of massless modes present in the 4d SCFT.

\subsection{Future Directions}

In this section, I briefly mention some open questions in the quantitative correspondence between the 4d gauge theories and 2d sigma models. Firstly, one may wonder if the correspondence stretches further than agreement between the spectra to agreement between correlation functions. Since the kinematics is obviously different in 2d and 4d, the relationship could not be exact. Nonetheless, one might wonder if, for example, the chiral rings of the two theories coincide. It would be interesting to explore this further.

\para
Another direction which merits further attention is the study of multiple, parallel, vortex strings. The moduli space ${\cal V}_{k,N}$ of $k$ vortices in $U(N_c)$ gauge theory with $N_f=N_c\equiv N$ has dimension,
\be {\rm dim}\,({\cal V}_{k,N})=2kN\ee
When the vortices are far separated, the moduli describe the positions and orientational modes of the $k$ strings. However, things become more murky as the vortices approach. A description of the $k$ vortex moduli space was presented in \cite{vib} in terms of a gauged linear model. One introduces an auxiliary $u(k)$ gauge field $u_\alpha$ on the mutual worldsheet of the strings. There are $N$ scalars, $\varphi_i$, transforming in the fundamental ${\bf k}$ representation, and a single complex scalar $Z$ transforming in the adjoint representation. The low-energy dynamics of $k$ vortices is then captured by the Lagrangian,
\be L_{k-vortex} = \Tr |{\cal D}_\alpha Z|^2+ |{\cal D}_\alpha\varphi_i|^2 - \Tr \lambda\, (\,[Z,Z^\dagger]+ \varphi_i\varphi_i^\dagger-r\,1_N)\label{multiv}\ee
where $\lambda$ is an adjoint-valued $u(k)$ Lagrange multiplier, $1_N$ is the unit $N\times N$ matrix and, as before, $r=4\pi/e^2$.

\para
The Lagrangian \eqn{multiv} is something of a caricature of the true vortex dynamics. After integrating out $\lambda$ we impose the $u(k)$-valued constraint $[Z,Z^\dagger]+\varphi_i\varphi_i^\dagger=r$. Subsequently dividing by $U(k)$ gauge transformations results in a sigma-model with target space of dimension $2kN$. The metric on this space does not coincide with the true metric on the vortex moduli space \cite{samols}. Nonetheless, it was argued in \cite{vib}, using a D-brane picture, that it correctly captures the topology and asymptotic behaviour of the moduli space. This was confirmed in \cite{jmatrix}, and the moduli space of two vortices in $U(2)$ gauge theory was studied in some detail in \cite{2vortex,2v2,2v3}. Thus, this non-Abelian 2d gauge theory should suffice to understand the dynamics of the vortices. In the absence of the adjoint field $Z$, the Lagrangian would describe a sigma-model with target space given by the Grassmannian $G(k,N)$. In this case, aspects of the theory were understood in \cite{phases,grass}. Introducing $Z$ renders the target space non-compact. It would be interesting to understand the low-energy dynamics of this theory and what it tells us about vortex interactions.

\para
Finally, there has recently been work looking at vortices in $U(1)\times SO(N)$ gauge theories, as well as other gauge groups \cite{otherg,otherg1,otherg2}. It should be possible to study the quantum dynamics of these vortex strings to see if they too reproduce the relevant Seiberg-Witten solution.

%

\section{Heterotic Vortex Strings and ${\cal N}=1$ Gauge Theories}

We have seen that in theories with ${\cal N}=2$ supersymmetry, the worldsheet of the vortex string captures much of the physics of the 4d theory in which it is embedded. In this section we turn to theories with ${\cal N}=1$ supersymmetry. We focus here on the theory with $N_f=N_c$, and vanishing quark masses $m_i=0$.

\para
BPS vortex strings once again preserve 1/2 of the supercharges, which now descend to  ${\cal N}=(0,2)$ supersymmetry on the worldsheet. They were dubbed ``heterotic vortex strings" in \cite{het}. The worldsheet theory includes a free ${\cal N}=(0,2)$ chiral mulitplet, containing the Goldstone mode $Z$ associated to the translation of the vortex, and a single right-moving Goldstino $\chi_+$ associated to the broken supersymmetry. The internal modes are now described by the ${\cal N}=(0,2)$ ${\bf CP}^{N-1}$ sigma-model \cite{het},
\be {\cal L}_{\rm vortex} &=& |{\cal D}_\alpha\varphi_i|^2 +
2i\left(\bar{\xi}_{+i}{\cal D}_-\xi_{+i} + \bar{\xi}_{-i}{\cal D}_+\xi_{-i}\right)
+ \lambda\left(|\varphi_i|^2-r\right)  + (\bar{\zeta}_-\bar{\xi}_{+i}\varphi_i+{\rm h.c.})\ . \ \ \ \ \ \ \ \ \ \ \ \  \label{v02}
\ee
This differs from the ${\cal N}=(2,2)$ theory of \eqn{v22} in two ways: firstly, the auxiliary field $\sigma$ is missing. Previously this induced a four-fermi coupling, but this is absent in the ${\cal N}=(0,2)$ theory. Secondly, while the right-moving fermions $\xi_+$ are constrained to live in the tangent bundle of ${\bf CP}^{N-1}$, there is no such constraint for the left-moving fermions $\xi_-$.

\para
However, there is an important caveat which must be borne in mind. With ${\cal N}=1$ supersymmetry, our theory does not have a mass gap when $v^2\neq 0$\ \footnote{We are interested here in the theory with $N_f=N_c$ and, as we review in Section \ref{n1mod}, the massless modes reflect the existence of a vacuum moduli space. This is different from the case of ${\cal N}=2$ supersymmetry, where the extra adjoint chiral multiplet $A$, together with the cubic superpotential ${\cal W}=\tilde{Q}AQ$, ensured that there were no massless modes when $v^2\neq 0$.}. This, in turn, means that there is a continuum of excitations of the vortex string, with energies reaching down to zero. How, then, to write down an effective action for the string? Which modes should we include? The Lagrangian above contains only the zero modes and none of the continuum. This is a standard approach to understand the ground states of solitons. For example, the same technique is used for the moduli space approximation for monopoles where the same issues arise \cite{manton}. Nonetheless, one should bear in mind that the validity of the low-energy effective action is more subtle that in the situation with mass gap and below we will use this low-energy theory only to understand the ground state of the vortex string. However, unlike the case of monopoles, there is yet another subtlety associated to this continuum of modes for the vortex string. The zero modes  arising from massless fields --- which, in the present case means the left-moving fermions $\xi_-$ --- are not normalizable. They suffer from a logarithmic infra-red divergence. This point has been stressed, for example, in \cite{shifhet1}. Nonetheless, it is imperative that these modes are included on the vortex worldsheet, for without them the theory would suffer from an anomaly. (In the gauged linear description, this is a standard gauge anomaly. In the non-linear sigma-model description, it is an anomaly of the type described in \cite{nonanom}). Here we proceed with the Lagrangian above, and treat these modes as if they had finite norm. As we shall see, the resulting physics does indeed correctly capture the dynamics of the 4d ${\cal N}=1$ SQCD. It would be desirable to have a better understanding of how to deal with these issues of non-normalizability.

\subsection{The ${\cal N}=(0,2)$ ${\bf CP}^{N-1}$ Sigma-Model}\label{02vor}

In this section we will examine the low-energy quantum dynamics of the model \eqn{v02}. While both the bosonic ${\bf CP}^{N-1}$ model and the ${\cal N}=(2,2)$ ${\bf CP}^{N-1}$ model have been studied in detail, less attention seems to have been paid to the ${\cal N}=(0,2)$ model.

\subsubsection*{The Spectrum}

Our first task is to describe the spectrum of the worldsheet theory \eqn{v02}. To put this in context, let us firstly recall the spectrum of the bosonic ${\bf CP}^{N-1}$ model, and of the ${\cal N}=(2,2)$ supersymmetric model.

\para
The bosonic ${\bf CP}^{N-1}$ model confines \cite{dadda,beaut}. This means that, although the elementary fields $\varphi_i$ live in the ${\bf N}$ of the global $SU(N)$ symmetry, the physical states form singlet and adjoint representations of $SU(N)$. The simplest way to see this is to start with the gauged linear model \eqn{cpn}. The $U(1)$ gauge field $u_\alpha$ here is merely an auxiliary construct, designed to impose the gauge equivalence $\varphi_i\equiv e^{i\alpha}\varphi_i$. However, in the quantum theory, the $U(1)$ photon becomes alive because, upon integrating out $\varphi_i$ in a $1/N$ expansion, a Maxwell kinetic term is generated,
\be {\cal L}_{\rm Maxwell} = \frac{1}{\Lambda^2_{2d}}\,G_{01}^2\ ,\label{mcpn}\ee
with $G_{01} = \partial_0u_1-\partial_1u_0$. (Note that, at one-loop, the running of the worldsheet coupling $r$ receives corrections only from the bosonic fields $\varphi_i$, and so $\Lambda_{2d}$ is independent of the fermion content). But, in two dimensions, the Coulomb force is linear. This means that charged particles, namely those associated with $\varphi_i$, are confined. Physical excitations are made from $\varphi$ and $\varphi^\dagger$ pairs and live in the adjoint and singlet representations of $SU(N)$.

\para
With ${\cal N}=(2,2)$ supersymmetry, the situation is different. The Maxwell term \eqn{mcpn} is once again generated, seemingly leading to confinement of $\varphi_i$. However, particles in the ${\bf N}$ representation of $SU(N)$ reappear in an interesting manner \cite{beaut}. The key to this is the existence of a $U(1)_A$ R-symmetry in the theory,
\be \xi_{\pm i}\rightarrow e^{\pm \alpha}\xi_{\pm i}\ .\label{u1a}\ee
There are now two important quantum effects associated to this symmetry. Firstly, it is afflicted by an anomaly, leaving only a discrete subgroup:  $U(1)_A\rightarrow {\bf Z}_{2N}$. Secondly, a fermionic condensate forms,
\be \langle \bar{\xi}_{+i}\xi_{-i}\rangle \sim \Lambda_{2d}\ .\label{cond}\ee
This last equation actually follows from the analysis of Section \ref{qworld} and, in particular, from \eqn{crut} and \eqn{whatsigma} when we set the masses to zero, $m_i=0$. The condensate spontaneously breaks the remnant discrete symmetry: ${\bf Z}_{2N}\rightarrow {\bf Z}_2$. This results in the $N$ isolated vacuum states that we saw in the previous section. The kinks which interpolate between these isolated vacua can be shown to transform in the ${\bf N}$ of the global $SU(N)$ \cite{beaut}. Thus the presence of the kinks means that the ${\bf CP}^{N-1}$ model with ${\cal N}=(2,2)$ supersymmetry does not confine.

\para
Having reviewed the standard ${\bf CP}^{N-1}$ lore, we now turn to the ${\cal N}=(0,2)$ model defined by the Lagrangian \eqn{v02}. This was studied in \cite{qhet}. There is once again a $U(1)_A$ symmetry \eqn{u1a} which is broken by an anomaly to ${\bf Z}_{2N}$. However, there can be no condensate \eqn{cond}. The reason is that the theory \eqn{v02} is invariant under a chiral global symmetry $SU(N)_L\times SU(N)_R$ which rotates $\xi_{-i}$ and $\xi_{+i}$ independently. A condensate \eqn{cond} would spontaneously break this continuous global symmetry, violating the Coleman-Mermin-Wagner theorem \cite{mw,col}. We conclude that the ${\cal N}=(0,2)$ ${\bf CP}^{N-1}$ model has a unique ground state. The confining force arising from the Maxwell term \eqn{mcpn} results in a spectrum of excitations which lie in the singlet, adjoint and bi-fundamental representations of the $SU(N)_L\times SU(N)_R$ global symmetry.

\para
In conclusion, the ${\cal N}=(0,2)$ sigma-model appears to be closer to the bosonic model since a condensate does not form, and there exists a single ground state. However, in a more important fashion, the ${\cal N}=(0,2)$ model differs from both its bosonic and ${\cal N}=(2,2)$ cousins: the theory does not have a mass gap. This fact follows immediately from the standard anomaly matching arguments of 't Hooft applied to the chiral $SU(N)_L\times SU(N)_R$ global symmetry. The low-energy spectrum must include massless fermions transforming in the bi-fundamental representation of $SU(N)_L\times SU(N)_R$. Moreover, there is also, at least, one further massless fermion which is a singlet of the global symmetry group. This is the Goldstino for, as we shall now show, the ${\cal N}=(0,2)$ model dynamically breaks supersymmetry \cite{qhet}.

\subsubsection*{Supersymmetry Breaking}

It is, in fact, a simple matter to see that the ${\cal N}=(0,2)$ ${\bf CP}^{N-1}$ model dynamically breaks supersymmetry. The smoking gun is a dynamically generated expectation value for the Lagrange multiplier $\lambda$. Since $\lambda$ is the auxiliary field in an ${\cal N}=(0,2)$ vector
multiplet (it is usually called $D$) and may be written as a supersymmetry variation $\delta
\zeta_-= \epsilon_- \lambda$, an expectation value for $\lambda$ means the theory dynamically breaks supersymmetry.
In fact, the expectation value for $\lambda$ was already
shown 30 years as a simple application of the $1/N$ expansion in
the bosonic sigma-model \cite{dadda,beaut}, and it is not
hard to check that the presence of fermions in the ${\cal N}=(0,2)$ model
do not change this conclusion\footnote{In contrast, supersymmetry
is not broken in the ${\cal N}=(2,2)$ supersymmetric model. This
can be traced to the presence of the extra bosonic field $\sigma$
which bears the burden of the expectation value instead of $\lambda$.}.
Ignoring the fermions, one first integrates out the bosonic fields
$\varphi_i$ to leave the partition function
\be Z_{\rm bose}=\int d\lambda\,dA_\mu\ \exp\left(-N\,\Tr\log
\left[-(\partial_\mu+iA_\mu)^2-\lambda\right]+i\int d^2x\ \lambda r
+\frac{\theta}{2\pi}\epsilon_{\mu\nu}\partial^\mu
A^\nu\right)\ .\nn\ee
The Lorentz invariant ground state has $A_\mu=0$, with $\lambda$ sitting at the
stationary point,
\be ir + N\int
\frac{d^2k}{(2\pi)^2}\,\frac{1}{k^2-\lambda+i\epsilon}=0\ .\ee
The integral can be performed exactly to give the supersymmetry breaking
expectation value,
\be \lambda = 4 \Lambda^2_{2d}\ ,\label{sgone}\ee
%

\subsubsection*{Another Way to See Supersymmetry Breaking}

There has been study of supersymmetry breaking in several other ${\cal N}=(0,2)$ models recently \cite{meng,shifhet2}. In this section, I briefly mention some unpublished work with Allan Adams and Brian Wecht which gives another perspective on supersymmetry breaking in the ${\cal N}=(0,2)$ ${\bf CP}^{N-1}$ model \cite{unpub}. This method uses the fact that the right-moving central charge $\hat{c}_R$ of any ${\cal N}=(0,2)$ superconformal fixed point is determined by the non-anomalous R-current $R$,
\be \hat{c}_R = \Tr\,R^2\ . \label{chat}\ee
Let's see what this gives for the central charge of our ${\cal N}=(0,2)$ ${\bf CP}^{N-1}$ model. The R-charges for the various fermions are $R(\xi_{+i})=R(\xi_{-i})=-1$ and $R(\zeta_-)=+1$. This results in the central charge $\hat{c}_R = N(-1)^2-N(-1)^2-1=-1$. But a negative central charge is nonsensical. What is going on? The point is that the equation \eqn{chat} assumes the existence of a ${\cal N}=(0,2)$ fixed point. A negative central charge is telling us that this assumption was incorrect, and can be seen as a diagnostic for supersymmetry breaking\footnote{As a check of this, we can look at the ${\bf CP}^{N-1}$ model with ${\cal N}=(2,2)$ supersymmetry. There is now one further Grassmann Lagrange multiplier with R-charge $R(\zeta_+)=+1$, so the resulting central charge is now $\hat{c}_R=0$, as befits a theory with a gap.}.

\subsection{Vortices and the Quantum Deformed Moduli Space}
\label{n1mod}

In the previous section, we studied the quantum dynamics of the heterotic vortex string. This semi-classical analysis is valid when $ev \gg \Lambda_{4d}$. We now turn to the study of vortices in the opposite regime, with $ev \ll \Lambda_{4d}$. Here, the 4d quantum effects are paramount. We will see how vortices behave in this regime, and interpret the worldsheet behaviour seen above, such as supersymmetry breaking, in terms of well-known facts about ${\cal N}=1$ SQCD.

\subsubsection*{The Vacuum Moduli Space}

Let us first review the dynamics of ${\cal N}=1$ SQCD with $N_f=N_c$ flavors \cite{seiberg}. We will firstly discuss the $SU(N_c)$ theory and only subsequently gauge baryon number $U(1)_B$ and introduce the FI parameter $v^2$.

\para
The classical theory has a moduli space of vacua parameterized by gauge invariant chiral operators, constructed from the quark chiral multiplets $Q_i$ and $\tilde{Q}_i$, with $i=1,\ldots,N$. The gauge invariant composites are the meson chiral superfield,
\be M_{ij}=\tilde{Q}_i Q_j\ee
together with a pair of baryon chiral superfields,
\be B=\epsilon_{a_1\ldots a_{N_c}}Q^{a_1}_1\ldots
Q^{a_{N_c}}_{N_c}\ \ \ \ \ \ ,\ \ \ \ \ \ \tilde{B}=\epsilon_{a_1\ldots
a_{N_c}}\tilde{Q}^{a_1}_1\ldots \tilde{Q}^{a_{N_c}}_{N_c}\ee
These are not independent. They obey the classical relationship
\be \det M - B\tilde{B}=0\label{clascons}\ee
The moduli space of vacua is given by the fields $M$, $B$ and
$\tilde{B}$ subject to the constraint \eqn{clascons}. Fluctuations along this  
space describe massless fields of the theory. The moduli space is singular at 
$\tilde{B}=B=0$
when ${\rm rank}(M)<N-2$. These singularities reflect the existence of new
massless gluons which emerge when the symmetry breaking is less
than maximal.

\para
The situation in the quantum theory is different. The classical
constraint \eqn{clascons} is corrected to \cite{seiberg},
\be \det M - B\tilde{B}=\Lambda_{4d}^{2N}\label{qcons}\ee
The manifold defined by \eqn{qcons} is smooth. The singularities of the classical
moduli space have been resolved, reflecting the confining nature
of the quantum theory.

\subsection*{Vortices in the Low-Energy Theory}

Let us now describe how vortices appear in the low-energy theory.
The $SU(N_c)$ theory does not have the topology to support vortex
strings. To introduce vortices we deform the theory by gauging the
$U(1)_B$ baryon symmetry. Of the low-energy fields, $M$ is neutral
under $U(1)_B$ while $B$ and $\tilde{B}$ have charge $+1$ and $-1$
respectively. We introduce a Fayet-Iliopoulos (FI) parameter
$v^2\ll\Lambda_{4d}^2$ for $U(1)_B$ which imposes the D-flatness condition on the
scalar fields,
\be |B|^2-|\tilde{B}|^2=v^2\label{d}\ee
Since $v^2>0$, we necessarily have $B\neq 0$ in vacuum. The
$U(1)_B$ invariant combination  $\tilde{B}B$ is then determined by
the meson expectation value through the constraint \eqn{clascons} or
\eqn{qcons}.

\para
The question that will concern us here is: when are the
vortex strings BPS? The equations describing a BPS string are the
first order Abelian vortex equations,
\be F_{12}^{B} = e^2 (|B|^2-|\tilde{B}|^2-v^2) \ \ \ \ ,\ \ \ \ {\cal
D}_1B=i{\cal D}_2B\ \ \ \ ,\ \ \ \  {\cal D}_1\tilde{B}=i{\cal
D}_2\tilde{B}\ \ \ \label{boglow}\ee
The key observation for our purposes is that these
equations have solutions only when $\tilde{B}=0$ \cite{penin,ddt,davis}.
This fact follows from a standard theorem in mathematics which states that there
exists no non-zero holomorphic section of negative degree --- see, for example,
\cite{phases}.

\para
What does this mean for the vortex theory? To make contact with the results of the previous section, we focus on the case $M=0$ for now. (This is identified as the vacuum with surviving $SU(N)_L\times SU(N)_R$ global symmetry. We will shortly relax this and look at vacua with $M\neq 0$). In the classical theory, the constraint \eqn{clascons} allows us to happily sit in the vacuum $|B|^2=v^2$, with $\tilde{B}=M=0$. Here BPS vortices exist; these are identified with the classical BPS vortices that we found in the previous section. However, in the quantum theory, things are rather different. The constraint \eqn{qcons} no longer allows $\tilde{B}=0$, and the vortex is no longer BPS. This is in agreement with our study of the microscopic vortex theory, in which supersymmetry was dynamically broken. We see that the breaking of supersymmetry on the worldsheet is the manifestation of the quantum deformation of the 4d moduli space.

\subsubsection*{The Spectrum}

In the case of ${\cal N}=2$ supersymmetry, we saw that the BPS spectrum of the worldsheet theory coincided with the BPS spectrum of the 4d theory (at $v^2=0$). With ${\cal N}=1$ supersymmetry, there are no BPS particles. Nonetheless, we shall see that there is still qualitative agreement between the quantum numbers of the 2d and 4d spectra.

\para
Let's start by providing an argument for why the spectra of the 2d and 4d theories should have anything to do with each other. Consider first the ${\cal N}=1$ $SU(N_c)$ theory. As reviewed above, the theory is believed to confine. The low-energy spectrum consists of a number of massless mesons and baryons, subject to the constraint \eqn{qcons}. The mesons transform in the bi-fundamental of the $F=SU(N)_L\times SU(N)_R$ flavor symmetry; the baryons are singlets. There are also a number of massive mesons, $Q^\dagger_iQ_j$ and $\tilde{Q}_i\tilde{Q}_j^\dagger$ transforming in the singlet and adjoint representations of $F$, together with massive baryons  transforming in a slew of tensor representations of $F$.

\para
We now deform the theory by gauging $U(1)_B$ and introducing the FI parameter $v^2$. For $v^2 \ll \Lambda_{4d}^2$, the baryons are now screened by the massive photon, while the mesons are left largely unaffected by this change. Some of these mesons may form weakly bound states with the vortex string. We can then ask what happens as we increase the ratio $v^2/\Lambda_{4d}^2$. This is a hard question. However, we may hypothesize that for $v^2\gg \Lambda_{4d}^2$, some of the mesons bound to the string remain light (i.e. of with mass of order $\Lambda_{4d}$). Such mesons would show up as internal excitations of the microscopic string theory.

\para
This hand-waving argument tells us to expect the spectrum of the vortex theory to contain some subset of the 4d meson spectrum. Let's now compare this expectation against our knowledge of the vortex spectrum. We saw in the previous section that the microscopic vortex theory has massless modes in the bi-fundamental representation of $F$, together with massive modes in the singlet and adjoint representations. This is in agreement with the meson spectrum of the 4d theory. The massive baryons in 4d are not seen in the worldsheet theory.

\subsubsection*{Worldsheet Supersymmetry Restoration}

There is another, more surprising, implication of the quantum deformed moduli space for the vortex string: supersymmetry restoration. Suppose that we sit in a classical vacuum with $\det M = \Lambda_{4d}^{2N}$. (Note that this is a strange thing to do: $\Lambda_{4d}$ is obviously a quantum generated scale. Here we are tuning a classical expectation value in anticipation of this scale). Then,  classically, we have $|\tilde{B}|^2 = \Lambda_{4d}^{4N}/v^2$, and no BPS vortices exist. However, quantum mechanically, we have $\tilde{B}=0$ and the vortex string regains its BPS status.

\para
Can we see this from the microscopic vortex theory? The answer is yes: we can write down a ${\cal N}=(0,2)$ theory which, classically, has no supersymmetric ground state. However, at one-loop, supersymmetry of the vacuum is restored. To see this, we first need to understand how the microscopic vortex theory \eqn{v02} is affected by turning on an expectation value for the meson field $M_{ij}$. The answer was given in \cite{qhet}: the vev $M$ induces a potential on the worldsheet of the vortex which preserves ${\cal N}=(0,2)$ supersymmetry
\be V = \frac{M_{ij}^\dagger M_{jk}}{v^2}\,\varphi_i^\dagger\varphi_k + \frac{M_{ij}}{v}\bar{\xi}_{i-}\xi_{j+}+{\rm h.c.}\label{02pot}\ee
This potential has a supersymmetric ground state, with $V=0$ only when ${\rm rank}(M)<N$. This is to be expected, for it coincides with the condition for the existence of classical BPS vortices. When ${\rm rank}(M)=N$, the ${\cal N}=(0,2)$ ${\bf CP}^{N-1}$ sigma model \eqn{v02}, together with the potential \eqn{02pot}, spontaneously breaks supersymmetry at the classical level.

\para
Let's now repeat our analysis of the quantum dynamics including the effects of this potential. Repeating the steps of Section \ref{02vor}, we integrate out the $\varphi_i$ fields to get an equation for the expectation value of $\lambda$. In the presence of the potential, equation \eqn{sgone} is replaced by
\be \det\left(\frac{M^\dagger M}{v^2}+\lambda\,1_N\right)=4^N\Lambda_{2d}^{2N}\ee
We see that, in agreement with the analysis of the quantum deformed moduli space, the theory has a supersymmetric ground state, with $\lambda=0$, provided that we tune the meson expectation values to,
\be \det M = 2^Nv^N\Lambda^N_{2d} \label{micro}\ee
To my knowledge, this is the first time that a supersymmetric vacuum is seen in a theory which, classically, breaks supersymmetry.

\para
It remains to match the 2d strong coupling scale $\Lambda_{2d}$ to the 4d scale $\Lambda_{4d}$. Recall that, for ${\cal N}=2$ supersymmetry, these coincided since the relationship $r=4\pi/e^2$ was preserved under RG flow \eqn{rrun}. With ${\cal N}=1$ supersymmetry, this is no longer the case: at one-loop, the beta-function for $r$ remains unchanged, while the beta-function for $e^2$ differs between the ${\cal N}=1$ and ${\cal N}=2$ theories. We instead match the $\Lambda$ parameters on
the worldsheet and in the bulk by first running $e^2$ down to the scale of the vortex tension, and subsequently running $r$ down to strong coupling, to find
\be \Lambda_{4d}^{2N} \sim v^N \Lambda_{2d}^N
\ee
We now see that, up to an overall constant that we haven't fixed, the microscopic vortex theory restores supersymmetry at the same point as predicted by the 4d quantum deformed moduli space: $\det M \sim \Lambda_{4d}^N$. However, it should be stressed that the above result holds at 1-loop in the 1/N expansion. It would be interesting to see if one can show that, for some finely-tuned $M$, there always exists a supersymmetric ground state within the class of theories given by (\ref{v02}) with potential (\ref{02pot}).

\subsection{Further Models}

As mentioned at the beginning of this section, there is a subtlety with the vortex worldsheet theory, arising from the fact that ${\cal N}=1$ SQCD does not have a mass gap. This results in a continuum of modes on the worldsheet, and the associated  logarithmic divergence of the left-moving fermions  $\xi_{i-}$. For this reason, a number of further models have been considered in the literature which do not suffer from this problem.

\para
The simplest such model is soft breaking of ${\cal N}=2$ theories to ${\cal N}=1$.  One may do this by simply adding a soft mass term $\mu$ for the adjoint chiral multiplet $A$. This led to a puzzle about the vortex dynamics since one might expect that this deformation in 4d would lead to a deformation on the worldsheet, in which the ${\cal N}=(2,2)$ supersymmetry of the ${\bf CP}^{N-1}$ sigma-model is broken softly to  ${\cal N}=(0,2)$ supersymmetry. The problem is that there exists no such deformation of the ${\cal N}=(2,2)$ sigma-model \cite{supersize}. This puzzle was resolved in \cite{het}, where it was pointed out that the worldsheet theory really has target space ${\bf C}\times {\bf CP}^{N-1}$. The ${\bf C}$ factor includes the overall translational mode and, in the ${\cal N}=(2,2)$ theory, two goldstino modes. It was shown in \cite{het,shifhet1} that there exists a unique deformation of the ${\bf C}\times {\bf CP}^{N-1}$ worldsheet theory, consistent with all the symmetries, which breaks ${\cal N}=(2,2)$ supersymmetry down to ${\cal N}=(0,2)$. The quantum dynamics of this model was studied in \cite{qhet,shifhet2}: with vanishing quark masses, $m_i=0$, the theory has $N$ degenerate vacua and breaks supersymmetry. For generic quark masses, $m_i\neq 0$, the degeneracy between the vacua is lifted.

\para
There is another, very interesting, model which alleviates the difficulty of the continuum of modes. This is the addition of a gauge singlet meson field $M$, transforming in the adjoint of the $SU(N_f)$ flavor group \cite{gsyagain}. Such fields are familiar from the Seiberg dual theories \cite{sdual}. This addition removes the difficulties with non-normalizable modes, without introducing further zero modes. However, it is not yet clear how, if at all,  the introduction of the meson field affects the interactions of the modes on the worldsheet.

\para
There is now a large literature studying the dynamics of vortices in different theories with ${\cal N}=1$ supersymmetry, much of it motivated by the ideas of dual confinement that were alluded to in Section 2.3. I refer the reader the original papers \cite{nonmon,abe,etoal,het,shifhet1,qhet,shifhet2,supersize,gsyagain, bolog,dkon,dynab} and the review articles \cite{syreview,konrev,kon1} for more details. To end this article, I would like to mention what, in my opinion, is an interesting open problem in this area: what does the vortex string know about Seiberg duality \cite{sdual}?  To answer this requires a study of  semi-local vortices in theories with $N_f>N_c$. There are tantalizing hints that ${\cal N}=(0,2)$ worldsheet theories may transform in an interesting fashion under the ubiquitous $N_f \rightarrow N_f-N_c$ transformation \cite{vib,sno} and various comments on the relationship to Seiberg duality have been made in \cite{eht1,soseiberg}. However, so far, there seems to be  no clean statement about how vortex dynamics is related to Seiberg duality and whether this connection implies the existence dual ${\cal N}=(0,2)$ worldsheet theories.

\section*{Acknowledgements}

My thanks to Amihay Hanany and Mohammad Edalati for enjoyable collaborations on
the work described in this review. I am also grateful to Allan Adams, Nick Dorey, Minoru Eto,
Cesar Gomez, Rajesh Gopakumar, Ken Konishi, David Kutasov, Muneto Nitta, Keisuke Ohashi, Norisuke Sakai, Misha Shifman, Walter Vinci, Brian Wecht and Alyosha Yung for interesting and valuable discussions over the years. This review is based on the ``Dottorato Lectures", given at Pisa in May 2008. I would like to thank the University of Pisa, GGI, Florence, and TIFR, Mumbai for hospitality during the completion of this work, and to INFN and ICTS for partial support. Finally, my thanks go to the Adams Prize committee for providing strong motivation to write this review. I am supported by the Royal Society.

\end{document}